\documentclass[useAMS,usenatbib,twocolumn]{mn2e}

\usepackage{amssymb,amsmath}
\usepackage{graphicx}
\usepackage{url}

\newcommand{\be}{\begin{equation}}
\newcommand{\ee}{\end{equation}}

\title[Magnetised filaments]{Polytropic models of filamentary interstellar clouds -- II \\
Helical magnetic fields} 

\author[Toci \& Galli]{Claudia Toci$^1$\thanks{E-mail: claudia@arcetri.astro.it}, 
Daniele Galli$^2$ \\
$^1$Dipartimento di Fisica e Astronomia, Universit\`a degli Studi di Firenze, Via G. Sansone 1, 
I-50019 Sesto Fiorentino, Italy \\
$^2$INAF-Osservatorio Astrofisico di Arcetri, Largo E. Fermi 5, I-50125 Firenze, Italy}

\begin{document}

\maketitle

\begin{abstract} 
We study the properties of magnetised cylindrical polytropes as
models for interstellar filamentary clouds, extending the analysis
presented in a companion paper (Toci \& Galli~2014a). We formulate
the general problem of magnetostatic equilibrium in the presence
of a helical magnetic field, with the aim of determining the degree
of support or compression resulting from the magnetisation of the
cloud. We derive scale-free solutions appropriate to describe the
properties of the envelopes of filaments at radii larger than the
flat-density region. In these solutions, the polytropic exponent
determines the radial profiles of the density and the magnetic
field.  The latter decreases with radius less steeply than the
density, and field lines are helices twisted over cylindrical
surfaces. A soft equation of state supports magnetic configurations
that preferentially compress and confine the filament, whereas in
the isothermal limit the field provides support.  For each value
of the polytropic exponent, the Lorentz force is directed outward
or inward depending on whether the pitch angle is below or above
some critical value which is a function of the polytropic exponent
only.
\end{abstract} 

\begin{keywords}
ISM: clouds -- magnetic fields. 
\end{keywords}

\section{Introduction}
\label{intro}

In a companion paper (Toci \& Galli~2014a, hereafter Paper~I) we
have analysed the structure and stability of unmagnetised cylindrical
polytropes, the simplest possible models of interstellar filaments, 
with the aim of interpreting the observations recently obtained
at submillimeter wavelengths by the {\em Herschel Space Observatory}
in a sample of nearby giant molecular clouds.
With respect to more complex simulations, polytropic models have
the advantage of requiring the minimum number of physical constants
(like the system's entropy $K$) and dimensionless parameters (like the
polytropic exponent $\gamma_{\rm p}=1+1/n$ and the adiabatic exponent $\gamma$). In
Paper~I, in analogy with previous studies of spherical clouds and cores,
we found that the observed radial density profiles of filaments are
well reproduced by a narrow range of $\gamma_{\rm p}$ ($1/3\lesssim \gamma_{\rm p}
\lesssim 2/3$) corresponding to negative values of the polytropic
index $n$. In particular, a good fit is obtained with $\gamma_{\rm p}\approx
1/2$ ($n\approx -2$), the polytropic exponent that characterises the
pressure of a superposition of low-amplitude undamped Alfv\'en waves.
It is of interest, therefore, to investigate further the properties of
magnetised filaments.

Starting from the work of Chandrasekhar \& Fermi~(1953), several models
have been proposed to study the structure and stability of filaments
threaded by poloidal and/or toroidal magnetic fields (Nagasawa~1987;
Fiege \& Pudritz~2000) or perpendicular to the main axis (Tomisaka et
al.~2014).  Unfortunately, there are few observational constraints on
the strength and/or morphology of the magnetic field within filaments.
Polarisation observations of background stars in the optical and near
infrared suggest that the field is generally uniform and perpendicular
to the filament, as, for example in the Serpens South cloud imaged
by Sugitani et al.~(2011) and the B211/B213/L1495 region in Taurus
(Palmeirim et al.~2013).  However, this is not a general rule. For
example, in the Taurus cloud, the field inferred by optical polarisation
is oriented mainly perperdicular to the main filaments B216 and B217
(Moneti et al.~1984; Goodman et al.~1992), whereas the L1506 filament
is almost parallel to the direction of the field (Goodman et
al.~1990). Indications of the presence of helical magnetic field twisted
along the filament's axis have been inferred in the dense core L1512 in
Taurus (Falgarone, Pety \& Phillips~2001; Hily-Blant et al.~2004) and in
NGC2024 in Orion~B (Matthews, Fiege \& Moriarty-Schieven~2002), but the
evidence is often indirect (see Gahm et al.~2006 and references therein
for a discussion of rotationally and magnetically twisted filaments).
A survey of filamentary molecular clouds in the Gould's Belt has shown
that these clouds tend to be oriented either parallel or perpendicular
to the ambient field direction in a bimodal fashion (Li et
al.~2013), a result that, if confirmed, indicates that the formation of
these structures is magnetically controlled.  It should be kept in mind,
however, that optical and near-infrared polarisation measurements do not
probe the magnetic field associated to the densest parts of the filaments
(Goodman et al.~1992).

Given the potentially important but still uncertain role played by the magnetic
field in determining the observable properties of filamentary clouds,
in this paper we examine the properties of magnetised polytropes
in cylindrical geometry, focusing in particular on the conditions 
for force balance in the radial direction.  
The models presented in this paper are magnetostatic by design. The 
underlying idea is that the evolution of real filaments can be analysed, as a 
first step, as a series of magnetostatic solutions as filaments accrete more 
material from the surrounding environment. As discussed in Paper~I, 
such a description does not necessarily imply zero a velocity field everywhere, as long
as the accretion speed becomes either subsonic or directed mostly parallel to the 
filament's axis in the central parts of the filament.

The paper is organised as follows: in
Sect.~\ref{equations} we derive the equations for the equilibrium
of a cylindrical cloud with a polytropic equation of state and a
magnetic field containing both a poloidal and toroidal component; in
Sect.~\ref{solutions} we solve these equations in some special  cases and
we generalise results found in previous studies; in Sect.~\ref{singular}
we present scale-free semi-analytical solutions applicable the power-law
envelopes of filamentary clouds; finally, in Sect.~\ref{conclusions}
we summarise our conclusions.

\section{Magnetized envelopes} 
\label{equations}

In a cylindrical system of coordinates, we consider the equilibrium structure of a self-gravitating
filament threaded by a helical magnetic field with poloidal and toroidal
components ${\bf B}_p$ and $B_\varphi$ defined in terms of the scalar
functions $\Phi(\varpi,z)$ and $\Psi(\varpi,z)$, respectively:
\be
{\bf B}_p=\nabla\times\left(\frac{\Phi}{2\pi\varpi}\, {\hat{\bf e}}_\varphi\right),
\qquad
B_\varphi=\frac{\Psi}{2\pi \varpi}.
\label{def_b}
\ee
Following Paper~I, we assume a polytropic equation of state,
\be
p=K\rho^{\gamma_{\rm p}},
\ee
where $K$ is a constant and $\gamma_{\rm p}=1+1/n$ is the polytropic exponent.
The equation of magnetostatic equilibrium is 
\be
-\nabla V+\frac{1}{\rho}\nabla p + {\bf F}_{\rm L}=0,
\label{force}
\ee
where $V$ is the gravitational
potential, related to the gas density $\rho$ by Poisson's equation
\be
\nabla^2 V=4\pi G\rho,
\label{pois}
\ee
and 
\be
{\bf F}_{\rm L}\equiv \frac{1}{4\pi\rho}(\nabla\times {\bf B})\times {\bf B}
\ee
is the Lorentz force per unit mass. With the definitions (\ref{def_b}), 
the Lorentz force becomes
\be
{\bf F}_{\rm L} =
-\frac{1}{16\pi^3\rho\varpi^2}[{\cal S}(\Phi)\nabla\Phi +
\Psi\nabla\Psi+\nabla\Phi\times\nabla\Psi]
\ee
where ${\cal S}$ is the Stokesian operator 
\be
{\cal S}(\Phi)=\frac{\partial^2\Phi}{\partial \varpi^2}+
\frac{\partial^2\Phi}{\partial z^2}-\frac{1}{\varpi}
\frac{\partial\Phi}{\partial\varpi}.
\ee
The condition of no Lorentz force in the azimuthal direction is
$\nabla\Phi\times\nabla\Psi=0$, which implies $\Psi=\Psi(\Phi)$. The
Lorentz force then reduces to
\be
{\bf F}_{\rm L}=-\frac{1}{16\pi^3\rho\varpi^2}
\left[{\cal S}(\Phi)+\Psi\frac{d\Psi}{d\Phi}\right]\nabla\Phi,
\label{fl_red}
\ee
This generalises the expression derived by Lizano \& Shu~(1989) in the
case of a poloidal field (see also Li \& Shu~1996 and Galli et al.~1999).
Taking the dot product of the force equation (\ref{force})
with ${\bf B}$ one obtains the condition of force balance along field
lines
\be
V+(1+n)K\rho^{1/n}=H(\Phi),
\label{along}
\ee
where $H(\Phi)$ is the Bernoulli constant.  The condition of force
balance across field lines (along $\nabla\Phi$) is
\be
-\frac{1}{16\pi^3\rho\varpi^2}\left[{\cal S}(\Phi)+\Psi\frac{d\Psi}{d\Phi}\right]=
\frac{dH}{d\Phi},
\label{across}
\ee
and Poisson's equation (\ref{pois}) then becomes
\begin{eqnarray}
\lefteqn{\frac{1}{\varpi}\frac{\partial}{\partial\varpi}
\left[\varpi\left(\frac{dH}{d\Phi}
\frac{\partial\Phi}{\partial\varpi}- 
\frac{1+n}{n}K\rho^{-1+1/n}
\frac{\partial\rho}{\partial\varpi}
\right)\right]+} \nonumber \\
& & \frac{\partial}{\partial z}\left(\frac{dH}{d\Phi}
\frac{\partial\Phi}{\partial z}-
\frac{1+n}{n}K\rho^{-1+1/n}
\frac{\partial\rho}{\partial z}\right)=4\pi G\rho.
\label{poisson}
\end{eqnarray}
Eq.~(\ref{fl_red}) and (\ref{across}) show that $H(\Phi)$ is a potential for 
the Lorentz force, $ {\bf F}_{\rm L}=\nabla H(\Phi)$.

The two coupled PDEs (\ref{across}) and (\ref{poisson}) are the two
fundamental equations of the problem.  To obtain a solution, one
must specify the two functions $H(\Phi)$ and $\Psi(\Phi)$, and apply
appropriate boundary conditions.  The arbitrariness in the choice of
the functional dependence of $H$ and $\Psi$ on the flux function $\Phi$
is a consequence of neglecting the dynamical evolution of the cloud.
The loss of information on the previous evolution of the cloud results
in the appearance of arbitrary functions that have to be determined from
physical considerations (see, e.g., Shu~1992).  Although in principle
any choice of $H$ and $\Psi$ is allowed, not all solutions would
lead to a physically meaningful model for a magnetised filamentary
cloud. Unfortunately the origin of filamentary clouds in the ISM is
not well understood (see discussion in Paper~I), and any attempt made
to reduce the arbitrariness in eq.~(\ref{across}) and (\ref{poisson})
must be necessarily {\em ad hoc}.

\subsection{Non-dimensional equations}

Eq.~(\ref{across}) and (\ref{poisson}) are generally valid under
azimuthal symmetry.  As a first simplification, we also assume cylindrical
symmetry ($\partial/\partial z=0$), reducing the problem to the solution
of a system of two coupled ordinary differential equations. 
We define a non-dimensional radius $\xi$ and density $\theta$ as in
Paper~I
\be
\varpi=\varpi_0\xi, \qquad \rho=\rho_c\theta^n,
\label{nondim1}
\ee
where
\be
\varpi_0=\left[\frac{\mp(1+n)K}{4\pi G\rho_c^{1-1/n}}\right]^{1/2}
\ee
is the radial scale length. Here, as in Paper~I, the subscripts ``$c$'' and ``$s$''
indicated quantities evaluated at the center and the surface of the filament,
respectively. We also define the non-dimensional magnetic
flux $\phi$, enthalpy $h$ and toroidal flux function
$\psi$ as
\be
\Phi=\left[\frac{\mp\pi (1+n)^3 K^3}{G^2\rho_c^{1-3/n}}\right]^{1/2}\phi,
\label{nondim2}
\ee
\be
\frac{dH}{d\Phi}=\left[\frac{\mp G^2\rho_c^{1-1/n}}{\pi (1+n) K
}\right]^{1/2}\frac{dh}{d\phi},
\label{nondim3}
\ee
\be
\Psi=\left[\frac{\mp 2\pi (1+n) K\rho_c^{1/n}}{G^{1/2}}\right] \psi,
\label{nondim4}
\ee
Here and in the following, as in Paper~I, the upper (lower) sign is
for $n\leq -1$ ($n>-1$), where $n$ is the polytropic index.

In nondimensional form, the components of the magnetic field are
\be
B_z=B_0b_z, \qquad B_\varphi=B_0b_\varphi,
\ee
where the scale factor $B_0$ is  
\be
B_0=[\mp 4\pi (1+n) p_c]^{1/2}.
\ee
The field components are then
\be
b_z=\frac{1}{\xi}\frac{d\phi}{d\xi}, \qquad
b_\varphi=\frac{\psi}{\xi}.
\ee
Similarly, the forces per unit volume acting on the system are: 
the pressure gradient,
\be
-\frac{1}{\rho}\nabla P=\pm F_0
\frac{d\theta}{d\xi}
\,{\hat {\bf e}_\varpi},
\ee
the Lorentz force
\be
{\bf F}_{\rm L}= F_0\frac{dh}{d\xi}\,{\hat{\bf e}_\varpi},
\label{fl}
\ee
and the gravitational force
\be
-\nabla V=-F_0\left(\frac{dh}{d\xi}\pm\frac{d\theta}{d\xi}\right)
\,{\hat{\bf  e}_\varpi},
\ee
where the scale factor $F_0$ is 
\be
F_0\equiv [\mp 4\pi (1+n)G p_c]^{1/2}=G^{1/2}B_0.
\label{def_f0}
\ee

With the definitions (\ref{nondim1})--(\ref{nondim4}), eq.~(\ref{across})
and (\ref{poisson}) can be written in non-dimensional form
\be
-\frac{1}{\xi^2\theta^n}\left[\left(\frac{d^2\phi}{d\xi^2}-\frac{1}{\xi}\frac{d\phi}{d\xi}\right)
\frac{d\phi}{d\xi}+\psi\frac{d\psi}{d\xi}\right]=
\frac{dh}{d\xi},
\label{across_n}
\ee
and
\be
\frac{1}{\xi}\frac{d}{d\xi}\left[\xi\left(\frac{dh}{d\xi}\pm\frac{d\theta}{d\xi}\right)\right]=\theta^n.
\label{along_n}
\ee
For the magnetic field to be well-behaved near the axis of the
cylinder, eq.~(\ref{across_n})--(\ref{along_n}) must be solved under
the boundary conditions $\theta(0)=1$ and $d\theta/d\xi(0)=0$, as in Paper~I,
plus the conditions $\xi^{-1}d\phi/d\xi\rightarrow {\rm const.}$
and $\xi^{-1}\psi\rightarrow 0$ for $\xi\rightarrow 0$, corresponding
to $B_z(0)={\rm const.}$ and $B_\varphi(0)=0$.

From eq.~(\ref{fl}) and (\ref{across_n}) one sees that in order to
provide support to the cloud an axial field must decrease with radius,
whereas a toroidal field must decrease with radius more rapidly
than $\varpi^{-1}$.  Thus in general the ability of a magnetic field
to support (or to compress) a filamentary cloud depends on the radial
profile of its strength and on the relative importance of the poloidal
and toroidal components.

\section{Special solutions}
\label{solutions}

\subsection{Force-free fields}

In general, magnetic force-free configurations have been applied to the
study of solar prominences, where the pressure gradients and self-gravity
of the plasma can be neglected with respect to the Lorentz force.
The possibility that force-free configurations can also arise in the
interstellar medium has been raised by Carlqvist, Kristen \& Gahm~(1998)
to explain the twisted appearance of some filamentary clouds on the
basis of the argument that only for a nearly force-free geometry the
electromagnetic effects are not too disruptive.

For $h=0$ the magnetic configuration is force-free, and the
equations for the density and the magnetic field are decoupled.
Eq.~(\ref{across_n}) becomes an equation only for the field,
\be
\left(\frac{d^2\phi}{d\xi^2}-\frac{1}{\xi}\frac{d\phi}{d\xi}\right)
\frac{d\phi}{d\xi}+\psi\frac{d\psi}{d\xi}=0, 
\label{forcefree} 
\ee
and eq.~(\ref{along_n}) reduces to the ordinary Lane-Emden
equation.  Several solutions of eq.~(\ref{forcefree}) are known. For
example, if $\psi=0$ (poloidal field) the only regular solution is
$\phi=A\xi^2$, with $A$ arbitrary constant, corresponding to the trivial
case of an uniform axial field.  If $\psi=k\phi$, with $k$ constant,
eq.~(\ref{forcefree}) is linear and reduces to Bessell's equation
with solution $\phi=C\xi J_1(k\xi)$, where $J_1(k\xi)$ is the Bessel
function of the first kind of order 1 and $C$ is a constant.  In this case
$b_z=CkJ_0(k\xi)$ and $b_\varphi=Ck J_1(k\xi)$. This is the Lundquist's
solution (Lundquist~1950). The field lines are helices that reverse
direction and handedness, since Bessell's functions are oscillatory.
The Lundquist solution is one of a class of solutions with oscillatory
behaviour that can be generated assuming a power-law dependence of
$\psi$ on $\phi$ and solving the resulting non-linear equation (Low \& Lou~1990).

In the following, we limit our analysis to the case $\phi>0$. Although
in general the flux function $\phi$ may change sign at one or more
radii, resulting in field reversals, the large-scale magnetic field
threading a molecular cloud core is expected to be the result of a
smooth distortion (i.e. without field reversals) of the relatively
uniform field characteristic of giant molecular clouds and galactic disks.

\subsection{Constant-$\beta$ solutions}

An axial field proportional to the square root of the gas pressure, such
that the plasma $\beta_z=(8\pi p/B_z^2)^{1/2}$ is spatially constant,
is a simple case to analyse because the total pressure is just a
scaled-up version of the gas pressure. In this case, eq.~(\ref{across_n})
and eq.~(\ref{along_n}) reduce to
\be
\frac{1+\beta_z^{-1}}{\xi}\frac{d}{d\xi}\left(
\xi\frac{d\theta}{d\xi}\right)=\pm\theta^n.
\label{tg}
\ee
With the scaling transformation $\xi\rightarrow (1+1/\beta_z)^{-1/2}\xi$,
this is the ordinary Lane-Emden equation for unmagnetized polytropic
cylinders.  Thus, the analysis of Paper~I remains valid, with the spatial
lenght scale $\varpi_0$ (and the core radius $\varpi_{\rm core}$)
increased by the factor $(1+1/\beta_z)^{1/2}$ (Talwar \& Gupta~1973,
Sood \& Singh~2004).  In particular, for all values of the polytropic
index $n$, the mass per unit length  and its critical or maximum value
determined in Paper~I are increased by the factor $1+1/\beta_z$.  The same
is not true in general for a purely toroidal field: only for an isothermal
equation of state a toroidal field with uniform $\beta_\varphi=(8\pi
p/B_\varphi^2)^{1/2}$ produces a simple rescaling of the
Lane-Emden equation (St\'odo\l kiewicz~1963). Therefore,
if $\gamma_{\rm p}=1$ the St\'odo\l kiewicz-Ostriker density
profile (see Paper~I) remains valid in the presence of a helical magnetic
field with constant $\beta_z$ and $\beta_\phi$.  These scaled isothermal
configurations have been studied by Nakamura, Hanawa \& Nakano~(1993).

\subsection{Fiege \& Pudritz's models}

Fiege \& Pudritz~(2000) solved the equations of magnetostatic equilibrium
for cylindrical clouds with an isothermal ($\gamma_{\rm p}=1$)
or logatropic ($\gamma_{\rm p}=0$) equation of state. They assumed a specific functional dependence
of the magnetic field strength on density and radius, namely
\be
B_z=\Gamma_z \rho, \qquad B_\varphi=\Gamma_\varphi\varpi\rho,
\label{fp}
\ee
with $\Gamma_z$, $\Gamma_\varphi$ being arbitrary constants.  This choice is
equivalent to the assumption that the ratio of magnetic flux to mass
per unit length is the same in all cylindrical shells of the filament
and that the toroidal component of the field is generated by a uniform
twisting of the filament through a fixed angle.  In our formalism,
eq.~(\ref{fp}) is equivalent to
\be 
\frac{d\Phi}{d\varpi}=2\pi\Gamma_z\varpi\rho,
\qquad \Psi=2\pi\Gamma_\varphi\varpi^2\rho.  
\ee 
Substituting these expressions in eq.~(\ref{across}) and (\ref{poisson}),
we obtain
\be 
-\frac{1}{4\pi}\left[\Gamma_z^2\frac{d\rho}{d\varpi}
+\Gamma_\varphi^2\frac{d}{d\varpi}(\varpi^2\rho)\right]=\frac{dH}{d\varpi},
\label{across_fp} 
\ee 
and 
\be
\frac{1}{\varpi}\frac{d}{d\varpi}\left[\varpi\left(\frac{dH}{d\varpi}
-\frac{1+n}{n}K\rho^{-1+1/n}\frac{d\rho}{d\varpi}\right)\right]=4\pi
G\rho.  
\label{poisson_fp} 
\ee 
Eq.~(\ref{across_fp}) can be integrated:
\be 
H=-\frac{1}{4\pi}\left(\Gamma_z^2\rho+\Gamma_\varphi^2
\varpi^2 \rho\right)+\mbox{const.}, \label{across_fp_1}
\ee 
and eq.~(\ref{poisson_fp}) can be written as 
\be
\frac{1}{\varpi}\frac{d}{d\varpi}\left\{\varpi\frac{d}{d\varpi}
\left[H-(1+n)K\rho^{1/n}\right]\right\}=4\pi G \rho.  
\label{poisson_fp_1} 
\ee 
Eq.~(\ref{across_fp_1}) and (\ref{poisson_fp_1}) generalise eq.~(D5) and
(D7) of Fiege \& Pudritz~(2000). The problem is reduced to the solution
of the second order ordinary differential equation~(\ref{poisson_fp_1})
for $\rho(\varpi)$, with $H$ given by eq.~(\ref{across_fp_1}).  These
generalised Fiege \& Pudritz~(2000) models have an asymptotic power-law
behaviour at large radii depending on the polytropic index $n$,
\be 
B_z \propto\rho\propto\varpi^\frac{2n}{1-n},
\qquad B_\varphi\propto\varpi\rho\propto\varpi^\frac{1+n}{1-n}.  
\ee 
Thus, the magnetic field becomes asymptotically dominated by the toroidal
component decreasing as a power-law with exponent between $-1$ and $0$.
Such toroidal field has the effect of compressing the cloud (as shown
in Sect.~2).

\section{Central cores and envelopes of magnetised filaments}
\label{singular}

\subsection{Series expansion for small radii}

On the filament's axis, symmetry requires the toroidal component of
the electric current to vanish, and the axial current to be finite.
This implies that, to the lowest order for $\xi\rightarrow 0$, both $\phi$
and $\psi$ must decrease at least like $\xi^2$. A series expansion
\be
\phi\approx\phi_2\xi^2+\phi_4\xi^4+\ldots,
\qquad
\psi\approx\psi_2\xi^2+\ldots,
\label{exp1}
\ee
gives
\be
b_z\approx b_{0z}+b_{2z}\xi^2+\ldots,
\qquad
b_\varphi\approx b_{1\varphi}\xi+\ldots
\label{exp2}
\ee
with $b_{0z}=2\phi_2$, $b_{2z}=4\phi_4$, $b_{1\varphi}=\psi_2$.
The substitution in eq.~(\ref{across_n}) and (\ref{along_n}) of the 
expansions (\ref{exp1}) and the expansion for $\theta$
\be
\theta\approx 1\pm\frac{1}{4}\xi^2+\ldots 
\label{exp_t}
\ee
gives the Lorentz force per unit mass near the axis,
\be
F_{\rm L}=-4F_0(b_{0z}b_{2z}+b_{1\varphi}^2)\xi+\ldots
\ee
where $F_0>0$ is given by eq.~(\ref{def_f0}).  In any realistic model
for filamentary clouds where the axial magnetic field decreases with
radius following the density, the product $b_{0z}b_{2z}$ is negative,
and the Lorentz force associated to the poloidal field is directed
outward, providing support to the cloud. Conversely, the Lorentz force
associated to the toroidal field is directed inward, squeezing the
cloud. The behaviour of the density near the axis is $\rho/\rho_{\rm
c}\approx 1-(\varpi/\varpi_{\rm core})^2+\ldots$, where
\be
\varpi_{\rm core}\approx\frac{2\varpi_0}{[\mp
n(1+4b_{0z}b_{2z}+4b_{1\varphi}^2)]^{1/2}}.
\label{rcore}
\ee
The core radius is therefore increased by the poloidal field ($b_{0z}b_{2z}<0$) 
and decreased
by the toroidal field ($b_{1\varphi}^2>0$).  This result can be used to constrain 
the strength and the morphology of the magnetic field near the filament's axis.
As shown in Paper~I, the core radius predicted
by unmagnetised polytropic models is 
\be
\varpi_{\rm core} \approx  0.047 
\left(\frac{\sigma_c}{0.26~\mbox{km~s$^{-1}$}}\right)
\left(\frac{n_c}{2\times 10^4~\mbox{cm$^{-3}$}}\right)^{-1/2}\, {\rm pc}.
\ee
For example, a toroidal field in
the core region such that $b_{1\varphi}^2 \approx 1$ 
(corresponding to $B_{\varphi,c} \approx \pi G^{1/2} \varpi\rho_c$ in physical 
units) would reduce the core radius by a factor of $\sim 2$. 
Field strengths of this order are invoked in the models of 
Fiege \& Pudritz~(2000). 

\subsection{Scale-free solutions for large radii}

\begin{figure}
\includegraphics[height=6.95cm,width=8.6cm]{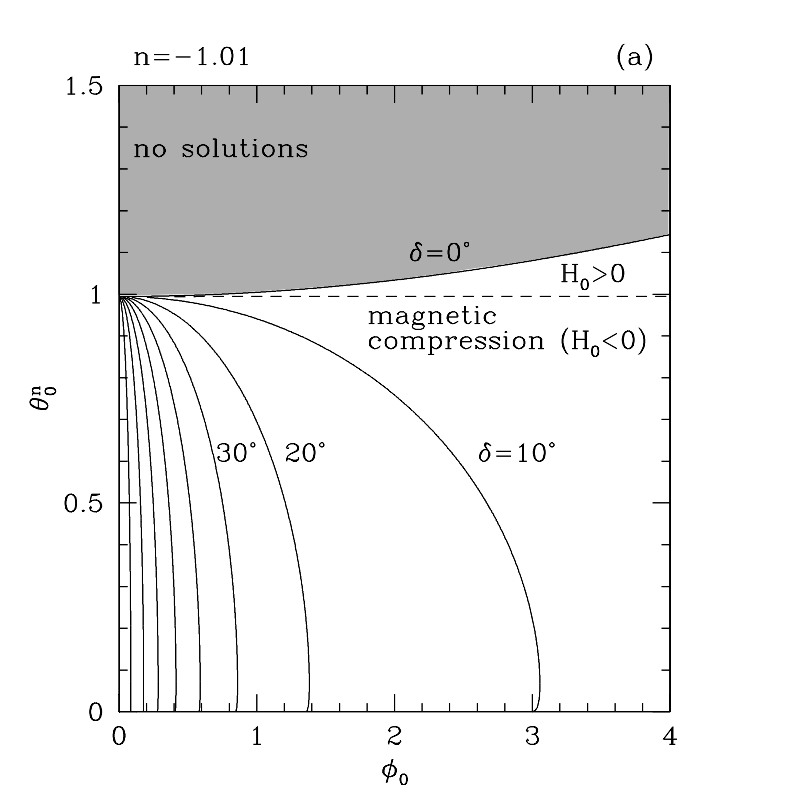}
\includegraphics[height=6.95cm,width=8.6cm]{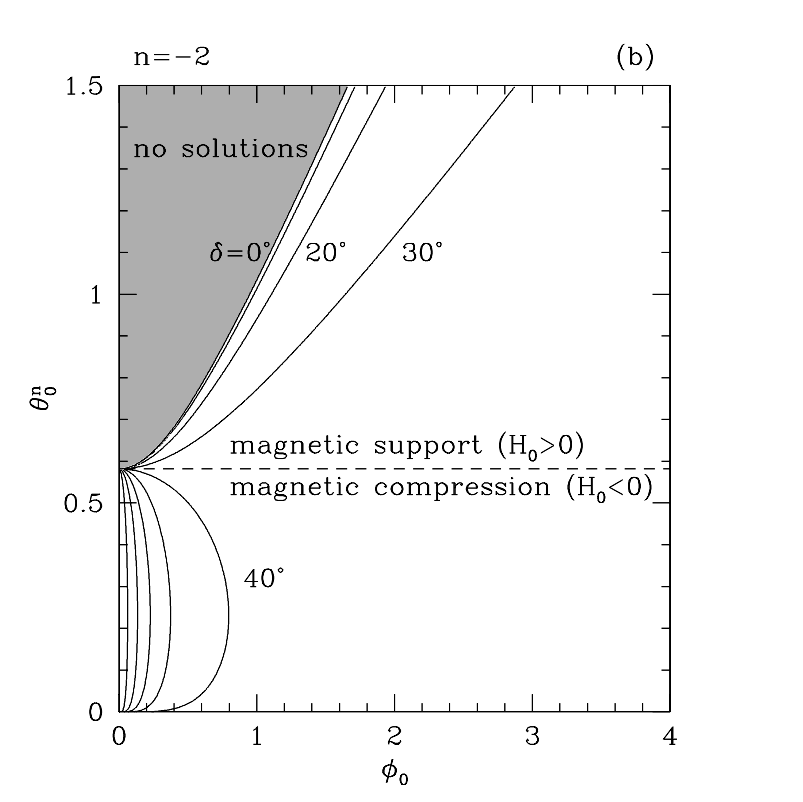}
\includegraphics[height=6.95cm,width=8.6cm]{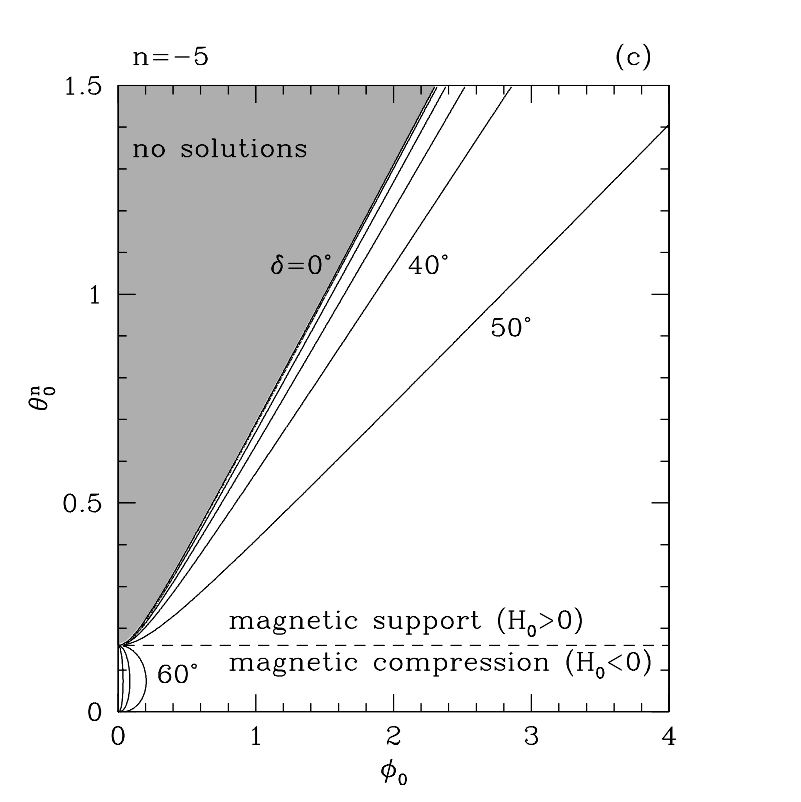}
\caption{Scale-free solutions in the $\phi_0$--$\theta_0^n$ plane for
$n=1.01$ (panel $a$), $n=-2$ (panel $b$) and $n=-5$ (panel $c$). 
The curves are for selected values of
the pitch angle $\delta$ from $0^\circ$ to $90^\circ$ in steps of $10^\circ$.
No solutions exist above the $\delta=0^\circ$ curve, representing models with
a pure poloidal field. No solutions exist in the {\em shaded} area above this curve. The 
{\em dashed}
line is for the pitch angle $\delta_{\rm ff}$, when the field becomes
force-free.  Above this line, the magnetic field provides support to the
cloud ($H_0>0$); below the dashed line, the magnetic field compresses
the cloud ($H_0<0$). }
\label{fig_sing} 
\end{figure}

To explore in the characteristics of polytropic magnetised filaments
at large radii (say, at radii much larger than the core radius given by
eq.~\ref{rcore}),  we seek asymptotic solutions of eq.~(\ref{across})
and (\ref{poisson}) without making specific assumption on the magnetic
field profile, except a power-law behaviour.  Recalling that a power-law
behaviour of the density in cylindrical polytropes is only possible for
$n\le -1$, in the following we restrict our analysis to this range of $n$.

For scale-free solutions, dimensional analysis requires $|{\bf B}| \propto
G^{1/2} \rho\varpi$. In addition, the enthalpy $H$ and the toroidal field
function $\Psi$ must have the following power-law dependence on $\Phi$:
\be
\frac{dH}{d\Phi}=H_0\left[\frac{G^2}{\pi |1+n|^n K^n}\right]^{\frac{1}{3-n}}
\Phi^{-\frac{1-n}{3-n}},
\ee
\be
\Psi\frac{d\Psi}{d\Phi}=4\alpha^2
\left[\frac{\pi^{2(2-n)}G^{1+n}}{(1+n)^{2n}K^{2n}}\right]^{\frac{1}{3-n}}
\Phi^{\frac{1+n}{3-n}},
\ee
where $H_0$ is a dimensionless constant that measures the deviation of
the poloidal field from force-free, $\alpha^2$ is a constant measuring
the strength of the toroidal field with respect to the poloidal field.
In non-dimensional form, with the definitions (\ref{nondim1})--(\ref{nondim4}), 
these expressions become
\be
\frac{dh}{d\phi}=H_0\phi^{-\frac{1-n}{3-n}},
\label{hsing}
\ee
and 
\be
\psi\frac{d\psi}{d\phi}=\alpha^2\phi^\frac{1-n}{3-n},
\label{psising}
\ee
Eq.~(\ref{hsing}) and (\ref{psising}) can be integrated to give
\be
h(\phi)=H_0\left(\frac{3-n}{2}\right)\phi^\frac{2}{3-n},
\ee
and
\be
\psi(\phi)=\pm\alpha\left(\frac{3-n}{2}\right)^{1/2}\phi^\frac{2}{3-n},
\ee
The system (\ref{across_n})--(\ref{along_n}) allows power-law solutions,
\be
\phi=\phi_0\,\xi^{\frac{3-n}{1-n}}, \qquad \theta^n=\theta_0^n\,\xi^{\frac{2n}{1-n}},
\ee
with the scale factors $\phi_0>0$ and $\theta_0>0$ given by 
\be
\frac{(3-n)(1+n)}{(1-n)^2}\phi_0^{\frac{2(2-n)}{3-n}}+\alpha^2\phi_0^{\frac{2}{3-n}}=
-H_0\theta_0^n,
\label{sing1}
\ee
\be
H_0\frac{2(3-n)}{(1-n)^2}\phi_0^{\frac{2}{3-n}}+\frac{4}{(1-n)^2}\theta_0=\theta_0^n.
\label{sing2}
\ee
In these scale-free models, all forces (gravity, pressure gradient, and
Lorentz force) decrease with radius with the same power-law slope. 
The Lorentz force is
\be
{\bf F}_{\rm L}=H_0\left(\frac{3-n}{1-n}\right)\phi_0^\frac{2}{3-n}\xi^\frac{1+n}{1-n} \hat{\bf e}_\varpi.
\ee 
and is directed inward for $H_0<0$ and outward for $H_0>0$.  For $H_0=0$, the magnetic
field is force-free. In this case, the density is given by the unmagnetised 
scale-free solution 
\be
\theta^n=\left[\frac{(1-n)^2}{4}\right]^{n/(1-n)}\xi^{2n/(1-n)},
\label{dens_unmag}
\ee
(see eq.~12 of Paper~I), and the components of the field are
\be
b_z=\phi_0\left(\frac{3-n}{1-n}\right)\xi^{\frac{1+n}{1-n}},
\label{bz}
\ee
\be
b_\varphi=\pm\alpha \phi_0^{\frac{2}{3-n}}
\left(\frac{3-n}{2}\right)^{1/2}\xi^\frac{1+n}{1-n}.
\label{bphi}
\ee
Thus, the magnetic field lines are helices twisted over cylindrical
flux tubes.  The magnetic field decreases with radius with a behaviour
intermediate between $\xi^{-1}$ for an isothermal equation
of state, and $\xi^0$ for a logatropic equation of state. For the
values of the polytropic index derived in Paper~I from fitting the radial
density profiles, $-3\lesssim n \lesssim -3/2$, the slope of the magnetic
field is in the range $-0.2$ to $-0.5$.  Thus, following
the discussion of Sect.~\ref{equations}, for all values of $n\le -1$
the axial field always supports the cloud and the toroidal field always
compresses it.  The net effect then depends on the relative strength of
the two components, determined by the value of the pitch angle $\delta$
(the angle between $B_z$ and $B_\varphi$)
\be 
\tan\delta=\frac{|B_\varphi|}{|B_z|}=
\frac{\alpha(1-n)}{[2(3-n)]^{1/2}}\phi_0^{-\frac{1-n}{3-n}}.
\ee 
For small $\delta$, the field is almost axial, and supports the envelope
with a Lorentz force directed outward. Increasing $\delta$ the toroidal
component becomes larger, squeezing the cloud with a Lorentz
force directed inward. At some particular $\delta_{\rm ff}$ the two
effects cancel out, and the field is force-free ($H_0=0$). This happens when
\be
\alpha=\alpha_{\rm ff}=\frac{[-(3-n)(1+n)]^{1/2}}{1-n}\phi_0^\frac{1-n}{1+n},
\ee
as can be obtained from eq.~(\ref{sing1}) setting $H_0=0$. Thus,
for each $n$, the field becomes force-free when the pitch angle is
\be
\tan\delta_{\rm ff}=\left(-\frac{1+n}{2}\right)^{1/2},
\label{dff}
\ee
a value that depends only on the polytropic index.  For $n=-1$ (logatropic case) the
force-free field is a poloidal field with uniform strength, while for
$n\rightarrow -\infty$ (isothermal case) the force-free field is toroidal and decreases as
$\varpi^{-1}$. For the values of $n$ derived in Paper~I for filamentary
cloud, $\delta_{\rm ff}$ varies between $26^\circ$ and $45^\circ$.

Fig.~\ref{fig_sing} shows the loci of solutions in the
$\phi_0$--$\theta_0^n$ plane (field strength vs. density in non-dimensional units), for three
values of the polytropic index: $n=-1.01$ (quasi-logatropic equation of
state), $n=-2$ (best-fit value for the observed filaments, see Paper~I) and $n=-5$
(quasi-isothermal equation of state). Each line is a locus of solutions with a fixed value of the
pitch angle $\delta$. The parameter space is divided
in two regions by the $H_0=0$ line of force-free configurations: for
$H_0>0$ the Lorentz force is directed outward and supports the cloud;
for $H_0<0$ the Lorentz force is directed inward and has the opposite
effect.  For $H_0=0$ the field is force-free, the pitch angle takes the value $\delta_{\rm ff}$
given by eq.~(\ref{dff}) and the density the value given by eq.~(\ref{dens_unmag}) independently on
the field strength $\phi_0$ (dashed lines in Fig.~\ref{fig_sing}). The line $\delta=0$ shows the locus of solutions
with a purely poloidal field. This line originates from the unmagnetised
solution $\theta_0^n=[(1-n)^2/4]^{n/(1-n)}$, $\phi_0=0$, and lies always
in the $H_0>0$ region of the diagram, indicating that a pure poloidal
magnetic field can only support, not compress, a cloud. The solutions
with $\delta=0$ are characterised by a density scale $\theta_0^n$
larger than the density scale of the unmagnetised model, due to the
extra support provided by the field. No solutions are possible above
the $\delta=0$ line. 

The curves in Fig.~\ref{fig_sing} show that for an increasing
field strength $\phi_0$, configurations with a fixed pitch
angle lower than the critical value $\delta_{\rm ff}$ support increasingly larger 
densities; whereas if the pitch angle is larger than $\delta_{\rm ff}$,
any increase in the field strength reduces the density that can be supported. However, in the 
latter case, equilibrium configuration only exists below a maximum value of $\phi_0$.
Thus, for a given field strength, 
there is always at least one solution with density larger than the unmagnetised 
solution, a pitch angle $\delta <\delta_{\rm ff}$ and magnetic effects dominated by the 
poloidal component (``magnetic support''), and
one (or two, or zero) solutions with density lower than the unmagnetised solution,
pitch angle $\delta>\delta_{\rm ff}$, and magnetic effects dominated by the toroidal component (``magnetic 
compression''). As shown by Fig.~\ref{fig_sing}, if the equation of state is soft (upper panel)
the largest fraction of parameter space is occupied 
by ``magnetic compression'' solutions; conversely, for a quasi-isothermal equation of state (lower panel)
``magnetic support'' solutions become dominant. This shows that the role of the magnetic field depends 
sensitively not only on the pitch angle but also on its dependence on radius via the polytropic exponent 
$\gamma_{\rm p}$.

The two cases $n=1.01$ and $n=-5$ in Fig.~\ref{fig_sing} illustrate the behaviour of the
solutions approaching the logatropic and the isothermal limit,
respectively. In the former case, solutions where the magnetic
field provides support progressively disappear. Already for a pitch
angle $\delta$ larger than about $4^\circ$, the hoop stresses of the
toroidal field dominate over the extra support of the dominant poloidal
field, squeezing the cloud.  Accordingly, the density is lower
than that of the unmagnetised solution. Increasing $-n$,
the region with $H_0<0$ shrinks, and the transition from support to
compression occurs at larger pitch angle (about $35^\circ$ for $n=-2$).
For $n=-5$ the allowed parameter space is largely populated by solutions
where the field provides support to the filament ($H_0>0$), but in
the limit $n\rightarrow -\infty$ the region of no solutions covers the
entire $\phi_0$--$\theta_0^n$ plane. The unmagnetised solution tends to
$\theta_0^n=0$ in this limit. In fact, even in the unmagnetised case,
no scale-free solution exists for an isothermal equation of state.

\subsection{Force between magnetised filaments}

The magnetic field discussed in the previous sections is generated
by electric currents flowing in the filaments. A toroidal field
$B_\varphi(\varpi)$, for instance, is associated to an electric current
$I(\varpi)\hat{\bf e}_z$ flowing along the filament given by
\be
I(\varpi)=\frac{c}{2}\int_0^\varpi [\nabla \times 
(B_\varphi {\hat{\bf e}}_\varphi)]
\varpi\, d\varpi=\frac{c}{2}\varpi B_\varphi(\varpi).
\label{curr}
\ee
Thus, two parallel filaments, separated by a distance $d$ larger than
their radii $\varpi_s$, behave as two electric wires and exert on each
other an electromagnetic force (per unit length) ${\bf F}_{em}$, repulsive
if the currents are aligned, attractive if they are anti-aligned, given by
\be
{\bf F}_{em}=\frac{2I_1I_2}{c^2 d}\, \hat{\bf e}_\varpi.
\ee
This electromagnetic force scales as the gravitational force (per unit length) 
${\bf F}_g$ between the two filaments,
\be
{\bf F}_g=\frac{2G\mu_1\mu_2}{d}\, \hat{\bf e}_\varpi,
\ee
where $\mu$ is the mass per unit length. Therefore 
their ratio is independent on the filaments separation
\be
{\cal R}\equiv \frac{|{\bf F}_{em}|}{|{\bf F}_g|}=\frac{I_1I_2}{c^2 G\mu_1\mu_2}
\sim \left(\frac{I}{cG^{1/2}\mu}\right)^2,
\ee
where the last approximation assumes that the two filaments have similar
properties.  According to eq.~(\ref{curr}), the electric current is zero
on the axis and equal to $c\varpi_s B_{\varphi,s}/2$ on the surface
(in our scale-free models for the filaments' envelopes, in which
$\varpi B_\varphi$ increases outwards,
the electric current is maximum at the surface). Taking the average,
we can set $I\approx c\varpi_s B_{\varphi,s}/4$, to obtain
\be
{\cal R}\approx 
\frac{1}{16}\left(\frac{\varpi_s B_{\varphi,s}}{G^{1/2}\mu}\right)^2.
\ee
Inserting numerical values,
\be
{\cal R}\approx 22 
\left(\frac{\varpi_s}{1~\mbox{pc}}\right)^2
\left(\frac{B_{\varphi,s}}{10~\mbox{$\mu$G}}\right)^2
\left(\frac{\mu}{10~\mbox{$M_\odot$~pc$^{-1}$}}\right)^{-2}.
\ee
The largest uncertainty on this result (in addition to the extreme idealisation of the picture) 
comes from the difficulty of estimating
the filament's outer radius $\varpi_s$ (see discussion in Arzoumanian
et al.~2011) and the corresponding value of the (toroidal) magnetic
field. Nevertheless the non-negligible numerical value of ${\cal R}$
indicates that the electromagnetic forces may play a role as
important as gravity in the interaction between magnetised filaments.

\section{Conclusions}
\label{conclusions}

We have derived general equations for magnetised filamentary clouds
with a polytropic equation of state, assuming cylindrical symmetry and
magnetostatic equilibrium.  The problem can be formulated in terms of two
partial differential equations for four unknowns: the density $\rho$,
the flux function $\Phi$ for the poloidal field, and two functions of
$\Phi$, namely the enthalpy $H(\Phi)$ and the toroidal field function
$\Psi(\Phi)$. 
Solutions can only be obtained by making additional assumptions, 
due to the lack of information on the previous evolution of
the system . These additional assumptions take usually the form of
specific choices of the dependence of the two component of the field on
density and/or radius (e.g. St\'odo\l kiewicz~1963, Talwar \& Gupta~1973,
Fiege \& Pudritz~2000).  In this work, we have avoided assumptions of this kind,
analysing the properties of the models in the region near the axis (the filament's
``core'') where the boundary conditions imposed by symmetry requirements
determine the behaviour of the solutions, and at large radii, where the
solutions are expected to approach a scale-free form for $n\le 1$. In this paper, we have explored
the range from a ``logatropic'' ($n=-1$, or $\gamma_{\rm p}=0$) to an ``isothermal''
($n\rightarrow -\infty$, or $\gamma_{\rm p}=1$) equation of state. In this
range, all variables have a power-law behaviour with profiles that
become flatter as the equation of state becomes softer. As the polytropic
exponent $\gamma_{\rm p}$ decreases from $1$ to $0$, the power-law exponents
range from $-2$ to $-1$ for the density and from $-1$ to $0$ for the
magnetic field.

Depending on the power-law slope, the magnetic field affects the radial
density profile of the cloud in opposite ways.  In the range $n\le
-1$, depending on the relative strength of the toroidal and poloidal
components, the magnetic field can either support or compress the
cloud (or be force-free). Pure poloidal fields, or with small toroidal
components (small pitch angle $\delta$) provide support to the cloud
($H_0>0$), allowing higher values of the envelope density $\theta_0^n$
than those resulting from thermal support alone. By
increasing the strength of the toroidal component ($\alpha^2>0$), the
effect of the field becomes extremely sensitive to the field strength
$\phi_0$: a small change in $\phi_0$ changes the sign of the Lorentz
force, from supporting to squeezing the cloud. In the latter case,
the density of the envelope is lower than the corresponding value in a
non magnetised filament. The confining effect of the field is enhanced
for softer equations of state, because the poloidal component becomes
increasingly uniform. In particular, for a logatropic equation of
states, all scale-free solutions are characterised by a Lorentz force
directed inward, and their density is lower than in the non magnetic
case. Conversely, all scale-free solutions converge to zero density in the
limit of an isothermal equation of state, because, as shown in Paper~I,
the ``natural'' asymptotic power-law behaviour of an isothermal cylinder
is reached only at infinite radius, where the density is zero.

{Within the limits of the idealised scale-free models for magnetised filaments
presented in this paper, our results suggest that a measure of the pitch angle 
of the magnetic field associated to filaments can provide a way to discriminate 
filaments that are compressed by the field from those that are supported. This 
can be accomplished by comparing the direction of the magnetic field around a
filament, as traced by optical/near-infrared polarisation, to the direction of the 
field within the filament, as traced by submillimetre polarisation. 
Preliminary results from the {\em Planck} satellite (Adam et al.~2014)
suggest that the field direction 
changes from the diffuse to the molecular gas, but the actual amount of field twisting
in a filament remains uncertain (see discussion in Sect.~\ref{intro}).
High-resolution observations of polarised dust emission 
inside the filaments (e.g. with ALMA) are needed to assess the role of large- and small-scale 
magnetic fields in their formation and evolution.

Finally, the magnetisation of interstellar filaments implies the presence of
electric currents flowing along and/or around them, producing attractive
(or repulsive) electromagnetic forces that enhance (or dilute) their
gravitational field and may affect their interactions. 

\section*{Acknowledgements}
It is a pleasure to acknowledge stimulating discussions with Philippe 
Andr\'e, Patrick Hennebelle and Evangelia Ntormousi. We also thank an anonymous referee for 
very useful comments that improved the presentation of the paper.


\begin{thebibliography}{}

\bibitem[Planck Collaboration et al.(2014)]{} 
Adam, R., Ade, P.~A.~R., et al.\ 2014, arXiv:1409.6728

\bibitem[Andr\'e et al.(2013)]{} 
Andr\'e, P., Di Francesco, J., Ward-Thompson, D., et al.\ 2013, in
Protostars \& Planets VI, arXiv:1312.6232

\bibitem[Arzoumanian et al.(2011)]{}
Arzoumanian, D., Andr\'e, P., Didelon, P., et al.\ 2011, A\&A, 529, L6

\bibitem[Carlqvist et al.(1998)]{} 
Carlqvist, P., Kristen, H., Gahm, G.~F.\ 1998, A\&A, 332, L5 

\bibitem[Chandrasekhar \& Fermi(1953)]{} 
Chandrasekhar, S., Fermi, E.\ 1953, ApJ, 118, 116 

\bibitem[Falgarone et al.(2001)]{}
Falgarone, E. Pety, J., Phillips, T.~G.\ 2001, ApJ, 555, 178

\bibitem[Fiege \& Pudritz(2000)]{} 
Fiege, J.~D., Pudritz, R.~E.\ 2000, MNRAS, 311, 85 

\bibitem[Gahm et al.(2006)]{}
Gahm, G.~F., Carlqvist, P., Johansson, L.~E.~B., Nikoli{\'c}, S.\ 2006, A\&A, 454, 201

\bibitem[Galli et al.(1999)]{} 
Galli, D., Lizano, S., Li, Z.~Y., Adams, F.~C., Shu, F.~H.\ 1999, ApJ, 521, 630 

\bibitem[Goodman et al.(1990)]{} 
Goodman, A.~A., Bastien, P., Menard, F., Myers, P.~C.\ 1990, ApJ, 359, 363 

\bibitem[Goodman et al.(1992)]{} 
Goodman, A.~A., Jones, T.~J., Lada, E.~A., Myers, P.~C.\ 1992, ApJ, 399, 108 

\bibitem[Hily-Blant et al.(2004)]{} 
Hily-Blant, P., Falgarone, E., Pineau Des For{\^e}ts, G., Phillips, T.~G.\ 2004, ApSS, 292, 285 

\bibitem[Li et al.(2013)]{} 
Li, H.-b., Fang, M., Henning, T., Kainulainen, J.\ 2013, MNRAS, 436, 3707 

\bibitem[Li \& Shu(1996)]{} 
Li, Z.-Y., Shu, F.~H.\ 1996, ApJ, 472, 211 

\bibitem[Lizano \& Shu(1989)]{} 
Lizano, S., Shu, F.~H.\ 1989, ApJ, 342, 834 

\bibitem[Low \& Lou(1990)]{}
Low, B.~C., Lou, Y.~Q.\ 1990,  ApJ, 352 

\bibitem[Lundquist(1950)]{}
Lundquist, S.\ 1950, Ark. Fys., 2, 361

\bibitem[Matthews et al.(2002)]{} 
Matthews, B.~C., Fiege, J.~D., Moriarty-Schieven, G.\ 2002, ApJ, 569, 304 

\bibitem[Moneti et al.(1984)]{} 
Moneti, A., Pipher, J.~L., Helfer, H.~L., McMillan, R.~S., Perry, M.~L.\ 1984, ApJ, 282, 508 

\bibitem[Nagasawa(1987)]{} 
Nagasawa, M.\ 1987, Progr. Theor. Phys., 77, 635 

\bibitem[Nakamura et al.(1993)]{} 
Nakamura, F., Hanawa, T., Nakano, T.\ 1993, PASJ, 45, 551 

\bibitem[Ostriker(1964)]{}
Ostriker, J.\ 1964a, ApJ, 140, 1056

\bibitem[Palmeirim et al.(2013)]{}
Palmeirim, P., Andr{\'e}, P., Kirk, J., et al.\  2013, A\&A, 550, A38 

\bibitem[Sood \& Singh(2004)]{} 
Sood, N.~K., Singh, K.\ 2004, Astrophys. Spa. Sci., 289, 55 

\bibitem[Stodolkiewicz(1963)]{}
St\'odo\l kiewicz, J.~S.\ 1963, Acta Astron. 13, 1

\bibitem[Shu(1992)]{} 
Shu, F.~H.\ 1992, The Physics of Astrophysics, Vol.~II, Gas Dynamics (Mill Valley: University Science Books), p.~320

\bibitem[Sugitani et al.(2011)]{} 
Sugitani, K., Nakamura, F., Watanabe, M., et al.\ 2011, ApJ, 734, 63 

\bibitem[Talwar \& Gupta(1973)]{} 
Talwar, S.~P., Gupta, A.~K.\ 1973, Astrophys. Sp. Sci., 23, 347 

\bibitem[Tomisaka(2014)]{} 
Tomisaka, K.\ 2014, ApJ, 785, 24 

\end{thebibliography}
\end{document}